\documentclass[conference]{IEEEtran}

\usepackage{cite}

\ifCLASSINFOpdf
  \usepackage[pdftex]{graphicx}
  \graphicspath{{../pdf/}{../jpeg/}}
  \DeclareGraphicsExtensions{.pdf,.jpeg,.png}
\else
  \usepackage[dvips]{graphicx}
  \graphicspath{{../eps/}}
   \DeclareGraphicsExtensions{.eps}
\fi
\usepackage[cmex10]{amsmath}
\usepackage{amssymb}

\begin{document}
%
\title{Scheduling and Pre-Conditioning in Multi-User \\MIMO TDD Systems}



%

\author{\IEEEauthorblockN{Jubin Jose\IEEEauthorrefmark{1},
Alexei Ashikhmin\IEEEauthorrefmark{2},
Phil Whiting\IEEEauthorrefmark{2},
Sriram Vishwanath\IEEEauthorrefmark{1}}
\IEEEauthorblockA{\IEEEauthorrefmark{1}Laboratory for Informatics, Networks and Communication (LINC)\\
Department of Electrical and Computer Engineering\\
The University of Texas at Austin, Austin, TX 78712}
\IEEEauthorblockA{\IEEEauthorrefmark{2}Bell Laboratories, Alcatel-Lucent Inc., Murray Hill, NJ 07974}}


\maketitle

\begin{abstract}
The downlink transmission in multi-user multiple-input multiple-output (MIMO) systems has been extensively studied from both communication-theoretic and information-theoretic perspectives. Most of these papers assume perfect/imperfect channel knowledge. In general, the problem of channel training and estimation is studied separately. However, in interference-limited communication systems with high mobility, this problem is tightly coupled with the problem of maximizing throughput of the system. In this paper, scheduling and pre-conditioning based schemes in the presence of reciprocal channel are considered to address this. In the case of homogeneous users, a scheduling scheme is proposed and an improved lower bound on the sum capacity is derived. The problem of choosing training sequence length to maximize net throughput of the system is studied. In the case of heterogeneous users, a modified pre-conditioning method is proposed and an optimized pre-conditioning matrix is derived. This method is combined with a scheduling scheme to further improve net achievable weighted-sum rate.

\end{abstract}


%
\IEEEpeerreviewmaketitle

\section{Introduction}
\label{intro}

\begin{figure}[!t]
\centering
\includegraphics[width=2.5in]{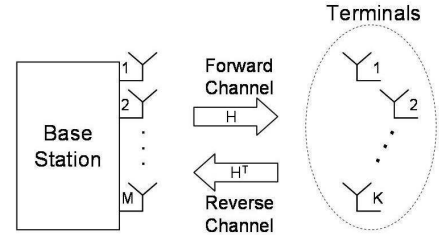}
\caption{Multi-User MIMO TDD System Model}
\label{figsysmodel}
\end{figure}

Downlink transmission in a multiple antenna setting is both a well studied and a complex problem with myriad parameters. A natural problem to be studied in this setting is to maximize throughput on the downlink while constraining the complexity at the terminals to be minimal. The problem of multi-antenna downlink transmission has been previously studied from many different perspectives \cite{Hochwald:space-time,Viswanath:sumcapacity,Goldsmith:fundamentallimits,Airy:multiuser_diversity}. In many of these papers, the channel is assumed to be known a-priori either perfectly or imperfectly at the base-station and/or terminals. The distinguishing feature of this paper is that we study the problem with no assumptions on channel knowledge both at the base-station and terminals (users). In addition, we consider very realistic and difficult communication regime when the forward SINRs are low ($\approx 0$ dB). We consider this regime since interference from neighboring base-stations does
not allow one to make SINRs larger.  Specifically, the scenario we study is the following: an $M$-element antenna array at the base-station, and single antennas at the $K(\leq M)$ autonomous terminals as shown in Fig. \ref{figsysmodel}. The channel is assumed to undergo block fading with a coherence interval of $T$ symbols. A time-division duplex (TDD) operation is considered. In a TDD system, the reverse channel and forward channel share a reciprocity relationship. Our system model is a generalization of the system model considered in \cite{Marzetta:howmuchtraining}. We look at the net impact of training, estimation, scheduling and pre-conditioning on the throughput of the system.
 
The use of multiple antennas instead of single antennas at the transmitter and receiver in a point-to-point communication system has been shown to greatly improve the capacity of the wireless channels \cite{Foschini:space-time,Telatar:capacity}. Later, the sum capacity of the multiple-input multiple-output (MIMO) Gaussian broadcast channel has been shown to be achieved by dirty paper coding (DPC) \cite{Caire:gaussianbc,Viswanath:sumcapacity,Vishwanath:duality}. Recently, it was shown that DPC actually characterizes the full capacity region of the MIMO Gaussian broadcast channel \cite{Weingarten:capacity}. In addition to the assumption that channel is perfectly known at the transmitter and the receivers, DPC scheme requires enormous computational power making it challenging to implement in practice. Motivated by this, many precoding and scheduling schemes have been proposed to obtain near-optimal performance with low complexity in certain scenarios \cite{Boccardi:precoding,Airy:transmit_precoding,Shen:low_complexity,Jagannathan:scheduling}. However, these schemes are not applicable to the scenario we consider.

We first look at the homogeneous users scenario, where all users have same forward and same reverse signal to interference-plus-noise ratios (SINRs), and obtain a rigorous lower bound on the sum capacity. The lower bound obtained in this paper is tighter than the lower bound given in \cite{Marzetta:howmuchtraining}. The improvement comes from the scheduling strategy used which is simple, and in fact even considerably reduces the computational complexity of pre-conditioning. In this context, we also study the problem of optimizing the training sequence length and the number of users to maximize net throughput of the system. Next, we look at the more general heterogeneous users scenario and study the problem of maximizing achievable weighted-sum rate. We propose a modified pre-conditioning method and obtain an optimized pre-conditioning matrix under $M$-large assumption. We combine this method with a simple scheduling strategy to take advantage of instantaneous channel variations.

We organize the remaining sections of this paper as follows. Section \ref{modeldesc} describes the system model and Section \ref{channelest} explains the reciprocal training used. Section \ref{homousers} and Section \ref{heterousers} describe the schemes proposed to increase achievable sum/weighted-sum rate in homogeneous and heterogeneous scenarios, respectively. We provide numerical results in Section \ref{simresults} and discuss our conclusions in Section \ref{concl}. 

\subsection{Notations}
In this paper, bold font variables denote vectors or matrices. All vectors are column vectors. $(\cdot)^{T}$, $(\cdot)^{*}$, $(\cdot)^{\dagger}$ and $\textit{tr}(\cdot)$ denote transpose, conjugate, Hermitian and trace, respectively. $\mathbb{E}[\cdot]$ and $\mathbf{var}\{\cdot\}$ stand for expectation and variance operations, respectively. $\textit{diag} \{\mathbf{a}\}$ stands for the $L \times L$ diagonal matrix with diagonal entries equal to the $L$ components of $\mathbf{a}$.

\section{Model Description}
\label{modeldesc}

The base-station with $M$ antennas communicates with the $K$ independent users on both forward and reverse links as shown in Fig. \ref{figsysmodel}. The forward channel is characterized by $K\times M$ propagation matrix $\mathbf{H}$. We assume independent Rayleigh fading channels, which remains constant over a duration of $T$ symbols called the coherence interval. The entries of the channel matrix $\mathbf{H}$ are independent and identically distributed (i.i.d.) zero-mean, circularly-symmetric complex Gaussian $CN(0,1)$ random variables. Our model incorporates frequency selectivity of fading by using orthogonal frequency-division multiplexing (OFDM). Note that the duration of the coherence interval in symbols is chosen for the OFDM sub-band. Due to reciprocity, we assume that the reverse channel at any instant is the transpose of the forward channel. 

Let the forward and reverse SINRs associated with $k^{th}$ user be $\rho_{fk}$ and $\rho_{rk}$, respectively. These forward and reverse SINRs remain fixed throughout the channel uses. On the forward link, the signal received by the $k^{th}$ user is
\begin{equation}
\label{forlink}
x_{fk}=\sqrt{\rho_{fk}}\:\mathbf{h}^{T}_k\mathbf{s_f}+w_{fk}
\end{equation}
where $\mathbf{h}^{T}_k$ is the $k^{th}$ row of the channel matrix $\mathbf{H}$ and $\mathbf{s_f}$ is the $M\times 1$ vector in which information symbols to be communicated are embedded. The components of the additive noise vector $[w_{f1} \: w_{f2}\:\cdots \:w_{fK}]$ are i.i.d. $CN(0,1)$. The average power constraint at the base-station during transmission is
$\mathbb{E}[\|\mathbf{s_f}\|^{2}]=1$
so that the total transmit power is fixed irrespective of its number of antennas. On the reverse link, the vector received at the base-station is
\begin{equation}
\label{revlink}
\mathbf{x_r}=\mathbf{H}^{T} \mathbf{E}_r  \mathbf{s_r}+\mathbf{w_r}
\end{equation}
where $\mathbf{s_r}$ is the signal-vector transmitted by the users and $\mathbf{E}_r = \textit{diag}\{[\sqrt{\rho_{r1}}\:\sqrt{\rho_{r2}}\:\cdots\:\sqrt{\rho_{rK}}]^{T}\}$. The components of the additive noise $\mathbf{w_r}$ are i.i.d. $CN(0,1)$. There is power constraint at every user during transmission given by
$\mathbb{E}[\|s_{rk}\|^{2}]= 1$
where $s_{rk}$ is the $k^{th}$ component of $\mathbf{s_r}$.

\section{Channel Estimation}
\label{channelest}

Channel reciprocity is one of the key advantages of TDD systems over frequency-division duplex (FDD) systems. We exploit this property to perform channel estimation by transmitting training sequences on the reverse link. Every user transmits a sequence of training signals of $\tau_{rp}$ symbols duration in every coherence interval. We assume that these training sequences are known a-priori to the base-station. The $k^{th}$ user transmits the training sequence vector $\sqrt{\tau_{rp}}\:\mathbf{\psi}^\dagger_k$. We use orthonormal sequences which implies $\mathbf{\psi}^\dagger_i \mathbf{\psi}_j = \delta_{ij}$ where $\delta_{ij}$ is the Kronecker delta. The use of orthogonal sequences restricts the maximum number of users to $\tau_{rp}$, i.e., $K\leq\tau_{rp}$. 

The corrupted training signals received at the base-station is
\begin{equation}
\label{}
\mathbf{Y}_r = \sqrt{\tau_{rp}}\:\mathbf{H}^{T}  \mathbf{E}_r \mathbf{\Psi^{\dagger}}+ \mathbf{V}_r
\end{equation}
where $\tau_{rp}\times K$ matrix $\mathbf{\Psi} = [\mathbf{\psi}_1 \, \mathbf{\psi}_2 \, \cdots \, \mathbf{\psi}_K]$ and the components of $M\times \tau_{rp}$ additive noise matrix $\mathbf{V}_r$ are i.i.d. $CN(0,1)$. The base-station obtains the LMMSE (linear minimum-mean-square-error) estimate of the channel
\begin{equation}
\label{chestimate}
\hat{\mathbf{H}}=\textit{diag}\left\{\left[\frac{\sqrt{\rho_{r1}\tau_{rp}}}{1+\rho_{r1}\tau_{rp}}\:\cdots \:\frac{\sqrt{\rho_{rK}\tau_{rp}}}{1+\rho_{rK}\tau_{rp}}\right]^T\right\}\mathbf{\Psi}^{T}\mathbf{Y}^T_r.
\end{equation}
This estimate $\hat{\mathbf{H}}$ is the conditional mean of $\mathbf{H}$ and hence, the MMSE estimate as well. By the properties of conditional mean and joint Gaussian distribution, the estimate $\hat{\mathbf{H}}$ is independent of the estimation error $\tilde{\mathbf{H}}=\mathbf{H}-\hat{\mathbf{H}}$. The components of $\hat{\mathbf{H}}$ are independent and the elements of its $k^{th}$ row are $CN\left(0,\frac{\rho_{rk}\tau_{rp}}{1+\rho_{rk}\tau_{rp}}\right)$. In addition, the components of $\tilde{\mathbf{H}}$ are independent and the elements of its $k^{th}$ row are $CN\left(0,\frac{1}{1+\rho_{rk}\tau_{rp}}\right)$.

\section{Homogeneous Users}
\label{homousers}

In this section, we focus on the special case where forward SINRs from the base-station to all users are equal and also reverse SINRs from all users to the base-station are equal, i.e., $\rho_{f1}=\cdots=\rho_{fK}=\rho_f$ and $\rho_{r1}=\cdots=\rho_{rK}=\rho_r$.

\subsection{Scheduling and Pre-Conditioning on Forward Link}
\label{schandprecond}

The base-station selects $N(\leq K)$ users among the $K$ users and pre-condition the information signals to be transmitted to these $N$ users. The scheduling strategy used to select the users is explained in Section \ref{schstrat}. Let the set of users selected be $S\subseteq \{1,2,\cdots,K\}$ with $N$ distinct entries. The base-station forms the $M\times 1$ transmission signal-vector $\mathbf{s}_f$, which drives the antennas, from the information symbol-vector $\mathbf{q}=[q_1 \: q_2 \:  \cdots \: q_N]^T$ for the selected users by pre-multiplying it with a pre-conditioning matrix. We use the pre-conditioning matrix
\begin{equation}
\label{matA}
\mathbf{A}_{S}=\frac{\mathbf{\hat{H}}^{\dagger}_{S}\left(\mathbf{\hat{H}}_{S}\mathbf{\hat{H}}^{\dagger}_{S}\right)^{-1}}{\sqrt{\textit{tr}\left[\left(\mathbf{\hat{H}}_{S}\mathbf{\hat{H}}^{\dagger}_{S}\right)^{-1}\right]}}
\end{equation}
which is proportional to the pseudo-inverse of the estimated channel. The $N\times M$ matrix $\mathbf{\hat{H}}_{S}$ is formed by the rows in set $S$ of matrix $\mathbf{\hat{H}}$. We use this pre-conditioning matrix because of the lack of any channel knowledge at the users. The pre-conditioning matrix is normalized so that
$\textit{tr}(\mathbf{A}_{S}^{\dagger}\mathbf{A}_{S})=1$.

The transmission signal-vector is given by
\begin{equation}
\label{trsignal}
\mathbf{s}_f=\mathbf{A}_{S}\mathbf{q}
\end{equation}
and the power constraint at the base-station is satisfied by imposing the condition
$\mathbb{E}[\|q_n\|^{2}]= 1, \forall n \in \{1,\cdots,N\}$.
From (\ref{forlink}) and (\ref{trsignal}), we obtain the signal-vector received at the selected users to be
\begin{equation}
\label{recvec}
\mathbf{x}_{f}=\sqrt{\rho_{f}}\:\mathbf{H}_{S}\mathbf{A}_{S}\mathbf{q}+\mathbf{w}_{f}
\end{equation}
where $\mathbf{H}_{S}$ is the matrix formed by the rows in set $S$ of the matrix $\mathbf{H}$.

\subsection{Lower Bound on Sum Capacity }
\label{lbonsumcap}

In this section, we obtain a lower bound on the sum capacity of the system under consideration. The approach is similar to that in \cite{Hassibi:howmuchtraining,Marzetta:howmuchtraining}. The lower bound holds for any scheduling strategy used at the base-station which selects a fixed number of users. Recall that the base-station performs channel estimation as described in Section \ref{channelest}.

\newtheorem{lemma}{Lemma}
\newtheorem{corollary}{Corollary}
\newtheorem{theorem}{Theorem}

\begin{theorem}
\label{indvlb}
For the system under consideration, every selected user can achieve a downlink rate during data transmission of at least
\begin{equation}
\label{indvlbeq}
C_{ind-lb}=\log_2\left(1+\frac{\rho_f\mathbb{E}^2\left[\chi\right]}{1+\rho_f\left(\frac{1}{1+\rho_r\tau_{rp}}+\mathbf{var}\{\chi\}\right)}\right)
\end{equation}
bits/transmission where $\chi$ is the scalar random variable given by
$\chi=\left(\textit{tr}\left[\left(\mathbf{\hat{H}}_{S}\mathbf{\hat{H}}^{\dagger}_{S}\right)^{-1}\right]\right)^{-\frac{1}{2}}.$
\end{theorem}

\begin{IEEEproof}
Let $\mathbf{\tilde{H}}_{S}$ be defined as the matrix formed by the rows in set $S$ of the matrix $\mathbf{\tilde{H}}$. The $N\times N$ effective forward channel matrix in (\ref{recvec}) is
\begin{IEEEeqnarray}{rCl}
\label{effchan}
\mathbf{G} &=& \sqrt{\rho_f}\:\mathbf{H}_{S}\mathbf{A}_{S}\\
&=& \sqrt{\rho_f}\left(\mathbf{\hat{H}}_{S}\mathbf{A}_{S} + \mathbf{\tilde{H}}_{S}\mathbf{A}_{S}\right)\nonumber\\
\label{Gmatrix}
&=& \sqrt{\rho_f}\left(\chi\mathbf{I}_N+\mathbf{\tilde{H}}_{S}\mathbf{A}_{S}\right).
\end{IEEEeqnarray}
From (\ref{effchan}) and (\ref{recvec}), we can write the signal received by the $n^{th}$ user as
\begin{equation}
\label{revsigatn}
x_{fn}=\mathbf{g}^{T}_{n}\mathbf{q}+w_{fn}
\end{equation}
where $\mathbf{g}^{T}_{n}$ is the $n^{th}$ row of $\mathbf{G}$. From (\ref{Gmatrix}), we obtain
\begin{equation}
\label{Grow}
\mathbf{g}^{T}_{n}=\sqrt{\rho_f}\left(\chi\mathbf{e}^T_n+\mathbf{\tilde{h}}^{T}_{S,n}\mathbf{A}_{S}\right)
\end{equation}
where $\mathbf{\tilde{h}}^{T}_{S,n}$ is the $n^{th}$ row of $\mathbf{\tilde{H}}_{S}$ and $\mathbf{e}^T_n$ is the $N\times 1$ vector with $n^{th}$ element equal to one and all other elements equal to zero. From (\ref{Grow}), we obtain
\begin{equation}
\label{gnmean}
\mathbb{E}\left[\mathbf{g}^{T}_{n}\right]=\sqrt{\rho_f}\:\mathbb{E}\left[\chi \right]\mathbf{e}^T_n
\end{equation}

and
\begin{equation}
\label{gnvar} \mathbb{E}\left[\mathbf{g}^{T}_{n}\mathbf{g}^{*}_{n}\right]=\rho_f\left(\mathbb{E}\left[\chi^2\right]+\frac{1}{1+\rho_r\tau_{rp}}\right).
\end{equation}


Adding $\mathbb{E}[\mathbf{g}^{T}_{n}]$ to and subtracting $\mathbb{E}[\mathbf{g}^{T}_{n}]$ from $\mathbf{g}^{T}_{n}$ in (\ref{revsigatn}), we obtain
\begin{IEEEeqnarray}{rCl}
x_{fn}&=&\mathbb{E}\left[\mathbf{g}^{T}_{n}\right]\mathbf{q}+\mathbf{\hat{g}}^{T}_{n}\mathbf{q}+w_{fn}\nonumber\\
\label{xsimplified}
&=&\sqrt{\rho_{f}}\:\mathbb{E}\left[\chi\right]q_n+\hat{w}_{fn}
\end{IEEEeqnarray}
where $\mathbf{\hat{g}}^{T}_{n}=\mathbf{g}^{T}_{n}-\mathbb{E}[\mathbf{g}^{T}_{n}]$ and $\hat{w}_{fn}$ is the zero-mean effective noise. Since the signal $\mathbf{q}$ is independent of $\mathbf{\hat{g}}^{T}_{n}$ and $\mathbb{E}[\mathbf{\hat{g}}^{T}_{n}]=0$, the signal $q_n$ is uncorrelated with the effective noise. Using (\ref{gnmean}) and (\ref{gnvar}), we obtain the variance 
\begin{IEEEeqnarray}{rCl}
\mathbf{var}\left\{\hat{w}_{fn}\right\}&=&\mathbb{E}\left[ \mathbf{\hat{g}}^{T}_{n} \mathbf{q}\mathbf{q}^{\dagger} \mathbf{\hat{g}}^{*}_{n}\right] +\mathbb{E}\left[\|w_{fn}\|^2\right]\nonumber\\
&=&\mathbb{E}\left[ \mathbf{\hat{g}}^{T}_{n} \mathbb{E}\left[\mathbf{q}\mathbf{q}^{\dagger}\right|\mathbf{g}_{n}] \mathbf{\hat{g}}^{*}_{n}\right] +\mathbb{E}\left[\|w_{fn}\|^2\right]\nonumber\\
&=&\mathbb{E}\left[\mathbf{g}^{T}_{n}\mathbf{g}^{*}_{n}\right] -\mathbb{E}\left[\mathbf{g}^{T}_{n}\right]\mathbb{E}\left[\mathbf{g}^{*}_{n}\right]
+\mathbb{E}\left[\|w_{fn}\|^2\right]\nonumber\\
\label{varianceterm}
&=&1+\rho_{f}\left(\frac{1}{1+\rho_r\tau_{rp}}+\mathbf{var}\left\{\chi\right\}\right).
\end{IEEEeqnarray}

Under the assumption that the users are aware of the scheduling strategy, $\mathbb{E}\left[\mathbf{\chi}\right]$ is known to the users. We obtain a lower bound on the downlink capacity of every selected user during data transmission by assuming worst-case noise distribution, which is uncorrelated Gaussian noise with same variance \cite{Hassibi:howmuchtraining}. Thus, from (\ref{xsimplified}) and (\ref{varianceterm}), we obtain (\ref{indvlbeq}) which completes the proof.
\end{IEEEproof}

\begin{corollary}
\label{sumlb}
For the system with homogeneous users considered, a lower bound on the sum capacity is
\begin{equation}
\label{csumlb}
C_{sum-lb}=\max_{N\leq K,\: N\in \mathbb{I}^+} N \cdot C_{ind-lb}.
\end{equation}
\end{corollary}

\subsection{Scheduling Strategy}
\label{schstrat}

The need for explicit scheduling arises due the use of pseudo-inverse based pre-conditioning of the information symbols. With perfect channel knowledge at the base-station ($\hat{\mathbf{H}}={\mathbf{H}}$) and no scheduling ($N=K$), the pseudo-inverse based pre-conditioning diagonalizes the effective forward channel and every user sees statistically identical effective channel irrespective of its actual channel. The inability to vary the effective gains to the users depending on their channel states is due to lack of any channel knowledge at the users. This possibly causes a reduction in achievable sum rate. Motivated by this, we propose a scheduling strategy which explicitly selects $N\leq K$ users before pre-conditioning.

In every coherence interval, the channel estimate at the base-station is used to select the $N$ users with largest estimated channel gains. Let $\mathbf{\hat{h}}_{(1)}^T,\mathbf{\hat{h}}_{(2)}^T,\cdots ,\mathbf{\hat{h}}_{(K)}^T$ be the norm-ordered rows of the estimated channel matrix $\mathbf{\hat{H}}$. Then, the matrix $\mathbf{\hat{H}}_{S}$ is given by
$\mathbf{\hat{H}}_{S}=[\mathbf{\hat{h}}_{(1)}\:\mathbf{\hat{h}}_{(2)}\:\cdots \:\mathbf{\hat{h}}_{(N)}]^T$
and the lower bound in (\ref{indvlbeq}) becomes
\begin{equation}
\label{maxsch}
C_{ind-lb}=\log_2\left(1+\frac{\rho_f\left(\frac{\rho_r\tau_{rp}}{1+\rho_r\tau_{rp}}\right)\mathbb{E}^2\left[\eta\right]}{1+\rho_f\left(\frac{1}{1+\rho_r\tau_{rp}}+\frac{\rho_r\tau_{rp}}{1+\rho_r\tau_{rp}}\mathbf{var}\{\eta\}\right)}\right).
\end{equation}
Here, the random variable
$\eta=\left(\textit{tr}\left[\left(\mathbf{U}\mathbf{U}^{\dagger}\right)^{-1}\right]\right)^{-\frac{1}{2}}$
where $\mathbf{U}$ is the $N\times M$ matrix formed by the $N$ rows with largest norms of a $K\times M$ random matrix $\mathbf{Z}$ whose elements are i.i.d. $CN(0,1)$. We provide numerical results showing the improvement obtained by using this strategy in Section \ref{simresults}.

\subsection{Net Achievable Sum Rate}

Net achievable sum rate accounts for the reduction in achievable sum rate due to training. In every coherence interval of $T$ symbols, first $\tau_{rp}$ symbols are used for training on reverse link, one symbol is used for computation (same assumption as in \cite{Marzetta:howmuchtraining}) and the remaining $T-\tau_{rp}-1$ symbols are used for transmitting information symbols. The number of users $K$ and the training length $\tau_{rp}$ can be chosen such that net throughput of the system is maximized. Thus, net achievable sum rate is defined as
\begin{equation}
\label{cnetach}
C_{net}(M,\rho_f,\rho_r)=\max_{K, \tau_{rp}}\frac{T-\tau_{rp}-1}{T}C_{sum-lb}(\cdot)
\end{equation}
subject to the constraints $\tau_{rp} \le T-2$ and $K \le \min(M,\tau_{rp})$. $C_{sum-lb}(\cdot)$ in (\ref{cnetach}) is given by (\ref{csumlb}).

\section{Heterogeneous Users}
\label{heterousers}

In this section, we consider the general setting described in Section \ref{modeldesc} with heterogeneous users. Moreover, we study the problem of maximizing achievable weighted-sum rate. The motivation behind this problem is that many algorithms implemented in layers above physical layer assign weights to each user depending on various factors. We assume that these weights are pre-determined and known. We propose a modified pre-conditioning method and derive an optimized pre-conditioning matrix under $M$-large assumption. We further combine this with a scheduling strategy to obtain an improved lower bound on the weighted-sum capacity.

\subsection{Modified Pre-Conditioning}

The base-station obtains the $M \times 1$ transmission signal-vector $\mathbf{s}_f$ by pre-multiplying the information symbols $\mathbf{q}=[q_1 \: q_2 \:  \cdots \: q_K]^T$ with a pre-conditioning matrix as explained in Section \ref{schandprecond}. We propose a modified pre-conditioning matrix given by
\begin{equation}
\label{modifiedA}
\mathbf{A}_D=\frac{\mathbf{\hat{H}}^{\dagger}_{D}\left(\mathbf{\hat{H}}_{D}\mathbf{\hat{H}}^{\dagger}_{D}\right)^{-1}}{\sqrt{\textit{tr}\left[\left(\mathbf{\hat{H}}_{D}\mathbf{\hat{H}}^{\dagger}_{D}\right)^{-1}\right]}}
\end{equation}
where $\mathbf{\hat{H}}_{D}=\mathbf{D \hat{H}}$ and $\mathbf{D}=\textit{diag}\left\{\left[p_1^{-\frac{1}{2}}\:p_2^{-\frac{1}{2}}\:\cdots \:p_K^{-\frac{1}{2}}\right]\right\}$. The choice of $\mathbf{D}$ is explained in Section \ref{optimizedA}. From (\ref{forlink}), we obtain the signal-vector received at the users
\begin{equation}
\label{recvechetero}
\mathbf{x}_{f}=\mathbf{E}_f  \mathbf{H}\mathbf{A}_D\mathbf{q}+\mathbf{w}_{f}
\end{equation}
where $\mathbf{E}_f=\textit{diag}\{[\sqrt{\rho_{f1}}\:\sqrt{\rho_{f2}}\: \cdots\:\sqrt{\rho_{fK}}]^{T}\}$.

\subsection{Lower Bound on Weighted-Sum Capacity}

In this section, we generalize the lower bound derived in Section \ref{lbonsumcap} to heterogeneous users and weighted-sum rate.

\begin{theorem}
For the system under consideration, a lower bound on the downlink weighted-sum capacity during transmission is given by $C_{wt-lb}$
\begin{equation}
\label{cwt}
=\sum\limits_{k=1}^{K} w_k\log_2\left(1+\frac{\rho_{fk}p_k\mathbb{E}^2\left[\phi_F\right]}{1+\rho_{fk}\left(\frac{1}{1+\rho_{rk}\tau_{rp}}+p_k\mathbf{var}\{\phi_F\}\right)}\right).
\end{equation}
Here, the random variable $\phi_F$ is given by
\begin{equation}
\label{phif}
\phi_F=\left(\textit{tr}\left[\left(\mathbf{F}\mathbf{Z}\mathbf{Z}^{\dagger}\mathbf{F}\right)^{-1}\right]\right)^{-\frac{1}{2}}
\end{equation}
where $\mathbf{F}=\mathbf{D} \cdot \textit{diag}\left\{\left[\sqrt{\frac{\rho_{r1}\tau_{rp}}{1+\rho_{r1}\tau_{rp}}}\:\cdots \:\sqrt{\frac{\rho_{rK}\tau_{rp}}{1+\rho_{rK}\tau_{rp}}}\right]^T\right\}$ and $\mathbf{Z}$ is the $K \times M$ random matrix whose elements are i.i.d. $CN(0,1)$.
\end{theorem}

\begin{IEEEproof}
The effective forward channel in (\ref{recvechetero}) is
\begin{IEEEeqnarray}{rCl}
\mathbf{G} &=& \mathbf{E}_f  \mathbf{H}\mathbf{A}_D\nonumber\\
&=& \mathbf{E}_f\left(\mathbf{D}^{-1}\mathbf{\hat{H}}_D\mathbf{A}_D + \mathbf{\tilde{H}}\mathbf{A}_D\right)\nonumber\\
\label{effchanhetero}
&=& \mathbf{E}_f \left( \phi_F\mathbf{D}^{-1}+\mathbf{\tilde{H}}\mathbf{A}_D\right).
\end{IEEEeqnarray}
The remaining steps in this proof are similar to those in the proof of Theorem \ref{indvlb} and hence, we skip it. 
\end{IEEEproof}

\subsection{M-large Asymptotics and Optimization of Pre-Conditioning Matrix}
\label{optimizedA}

We wish to choose the matrix $\mathbf{D}$ such that $C_{wt-lb}$ in (\ref{cwt}) is maximized. However, this problem is hard to analyze. We consider the asymptotic regime $M/K \gg 1$. Apart from making the problem mathematically tractable, this asymptotic regime is interesting due to the following two reasons. i) In our system model, we observe that extra base-station antennas are always beneficial from numerical results given in Section \ref{simresults}. This observation was first made for homogeneous users in \cite{Marzetta:howmuchtraining}. ii) The system imposed constraints $K \leq \tau_{rp}$ and  $\tau_{rp}\leq T$ restrict the value of $K$.

It is known that
$\lim_{M/K\rightarrow \infty}\mathbf{Z}\mathbf{Z}^{\dagger} \rightarrow M\mathbf{I}_K$
where $\mathbf{Z}$ is the $K\times M$ random matrix whose elements are i.i.d. $CN(0,1)$. Therefore, under M-large approximation, random variable $\phi_F$ in (\ref{phif}) can be approximated to
\begin{equation}
\label{phifapprox}
\phi_F\approx \sqrt{\frac{M}{\textit{tr}\left(\mathbf{F}^{-2}\right)}}
\end{equation}
which is a constant. Substituting (\ref{phifapprox}) in (\ref{cwt}), we get
\begin{equation}
C_{wt-lb}\approx J(\mathbf{p})=\sum\limits_{i=1}^{K} w_i\log_2\left(1+\frac{\beta_ip_i}{\sum\limits_{j=1}^{K}\alpha_jp_j}\right)
\end{equation}
where $\alpha_j = (\frac{\rho_{rj}\tau_{rp}}{1+\rho_{rj}\tau_{rp}})^{-1}$ and $\beta_i =\frac{M\rho_{fi}}{1+{\rho_{fi}}({1+\rho_{ri}\tau_{rp}})^{-1}}$.

\begin{theorem}
Let $\mathbf{p}=[p_1\:p_2\:\cdots\:p_K]^T$ be any vector of non-negative real numbers and
$\mathbf{p}^{*}=\arg \max_{\mathbf{p}} J(\mathbf{p})$
then the set of possible values of $\mathbf{p}^{*}=c\mathbf{\overline{p}}^{*}$ where $c$ is any positive real number and $\mathbf{\overline{p}}^{*}=[\overline{p}_1^{*}\:\overline{p}_2^{*}\:\cdots\:\overline{p}_K^{*}]^T$ such that
\begin{equation}
\label{optimalpi}
\overline{p}_i^{*}=\left(\frac{w_i}{\lambda^{*}\alpha_i}-\frac{1}{\beta_i}\right)^{+}.
\end{equation}
The positive real number $\lambda^{*}$ is chosen such that the constraint $\sum\limits_{i=1}^{K}\alpha_i\overline{p}_i^*=1$ is satisfied.
\end{theorem}
\begin{IEEEproof}
We use Lagrange multipliers to obtain this result. Due to lack of space, we do not include the proof here.
\end{IEEEproof}

The optimized $\mathbf{\overline{p}}^{*}$ given by ({\ref{optimalpi}}) is substituted in (\ref{modifiedA}) to obtain the optimized pre-conditioning matrix. We use this optimized pre-conditioning matrix even when $K$ is comparable to $M$.

\subsection{Scheduling Strategy}

In our system model, the optimized values $\mathbf{\overline{p}}^{*}$ cannot depend on the instantaneous channel as no channel information is available to the users. Hence, we need explicit selection of users to take advantage of the instantaneous channel variations. In this section, we propose a scheduling strategy for heterogeneous users.

Let $\mathbf{z}_{1}^T,\mathbf{z}_{2}^T,\cdots ,\mathbf{z}_{K}^T$ be the rows of the matrix
\begin{equation}
\mathbf{Z}=\textit{diag}\left\{\left[\sqrt{\frac{1+\rho_{r1}\tau_{rp}}{\rho_{r1}\tau_{rp}}}\:\cdots \:\sqrt{\frac{1+\rho_{rK}\tau_{rp}}{\rho_{rK}\tau_{rp}}}\;\right]^T\right\}\mathbf{\hat{H}}
\end{equation}
where $\mathbf{\hat{H}}$ is the estimated channel given by (\ref{chestimate}). In every coherence interval, the users are ordered such that $\overline{p}_{(1)}^{*}\|\mathbf{z}_{(1)}^T\|^2 \geq \overline{p}_{(2)}^{*}\|\mathbf{z}_{(2)}^T\|^2 \geq \cdots \geq \overline{p}_{(K)}^{*}\|\mathbf{z}_{(K)}^T\|^2$ and information symbols are transmitted to the first $N$ users using the pre-conditioning matrix formed by the appropriate rows of the optimized pre-conditioning matrix as described in Section \ref{optimizedA}. The value of $N$ is chosen in order to maximize achievable weighted-sum rate. We denote this lower bound on achievable weighted-sum rate with scheduling by $C_{wt-lb}^{sh}(\cdot)$.

\subsection{Net Achievable Weighted-Sum Rate}

We define net achievable weighted-sum rate as
\begin{equation}
C_{wt-net}(M,K,\rho_f,\rho_r)=\max_{\tau_{rp}}\frac{T-\tau_{rp}-1}{T}C_{wt-lb}^{sh}(\cdot)
\end{equation}
subject to the constraints $\tau_{rp} \ge K$ and $\tau_{rp} \le T-2$.

\section{Numerical Results}
\label{simresults}
We provide numerical results in both homogeneous and heterogeneous users scenarios to show the performance benefits obtained using the various proposed schemes. We are interested in the realistic communication regime when forward and reverse SINRs are low. We consider this regime since interference from neighboring base-stations force systems to operate in this regime. Moreover, we are interested in high mobility users. Hence, we choose the system parameters for these scenarios.

\subsection{Homogeneous Users}
\begin{figure}[!t]
\centering
\includegraphics[width=88mm]{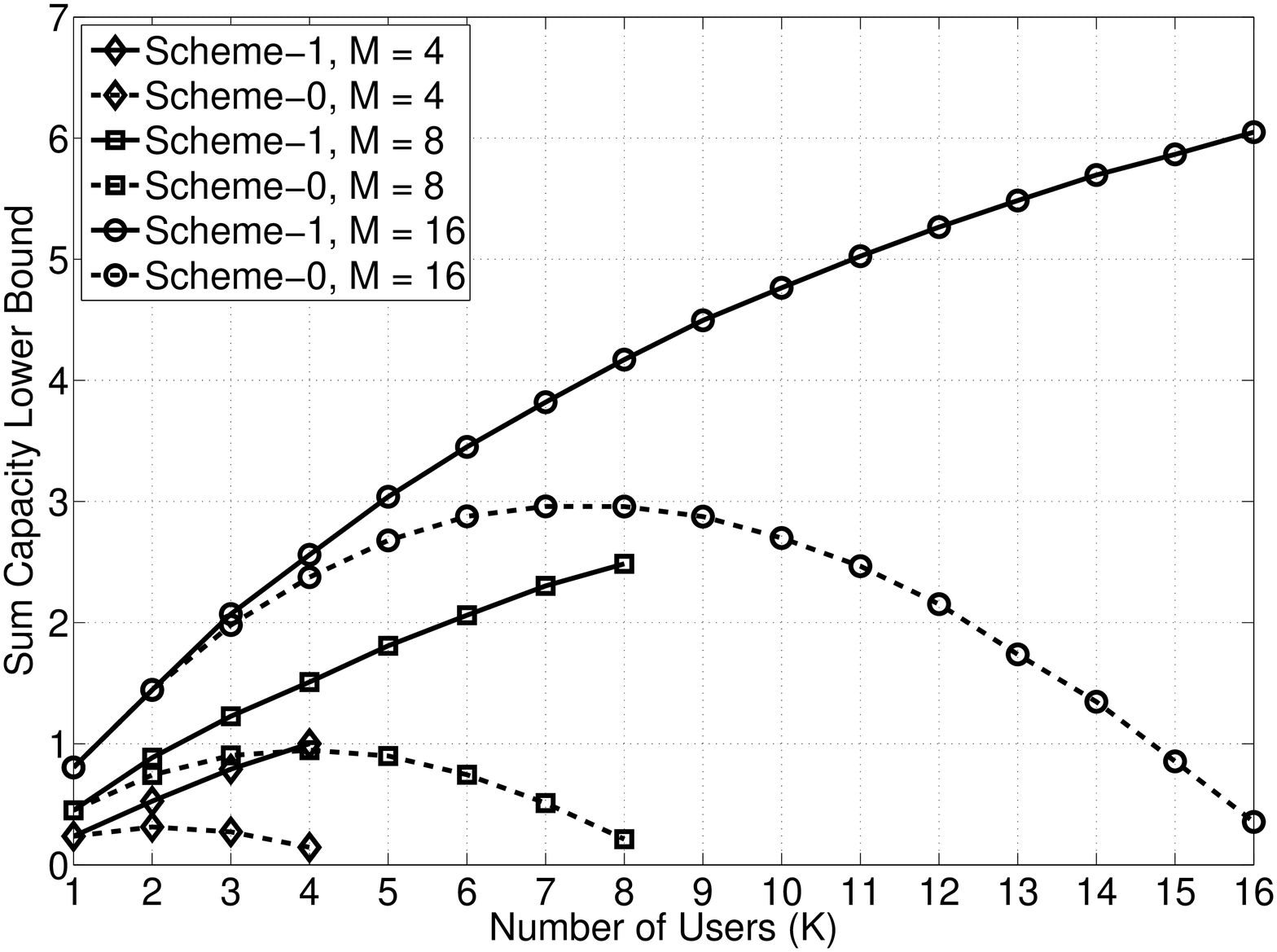}
\caption{Lower bound on the sum capacity with scheduling (Scheme-1) and without scheduling (Scheme-0)}
\label{fig_sim1}
\end{figure}

\begin{figure}[!t]
\centering
\includegraphics[width=88mm]{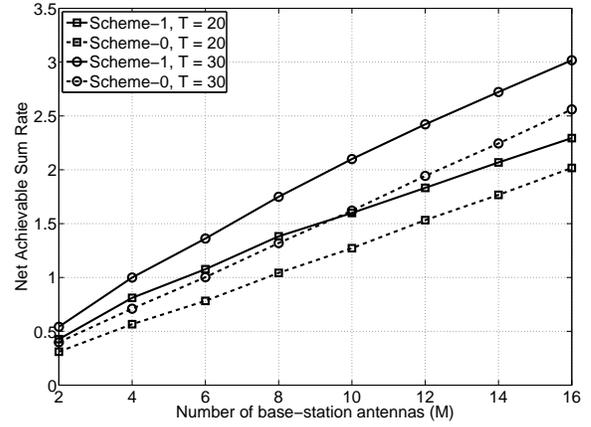}
\caption{Net achievable sum rate with scheduling (Scheme-1) and without scheduling (Scheme-0)}
\label{fig_sim2}
\end{figure}

We consider forward SINR $\rho_f$ of $0$ dB and reverse SINR $\rho_r$ of $-10$ dB. First, we keep the training sequence length equal to the number of users, i.e., $\tau_{rp}=K$. In Fig. \ref{fig_sim1}, we plot lower bound on the sum capacity with scheduling (Scheme-1) and without scheduling (Scheme-0) for $M=\{4,8,16\}$ and $K=\{1,2,\cdots,M\}$. Note that Scheme-0 is the lower bound obtained in \cite{Marzetta:howmuchtraining}. The proposed scheme gives significant improvement which implies that the scheme is capable of performing opportunistic scheduling. Next, in Fig. \ref{fig_sim2}, we plot net achievable sum rate versus $M$ for $T=\{20,30\}$. We observe that the net achievable sum rate increases with $M$ for all schemes. As expected, the proposed scheduling scheme (Scheme-1) outperforms Scheme-0. 

\begin{figure}[!t]
\centering
\includegraphics[width=88mm]{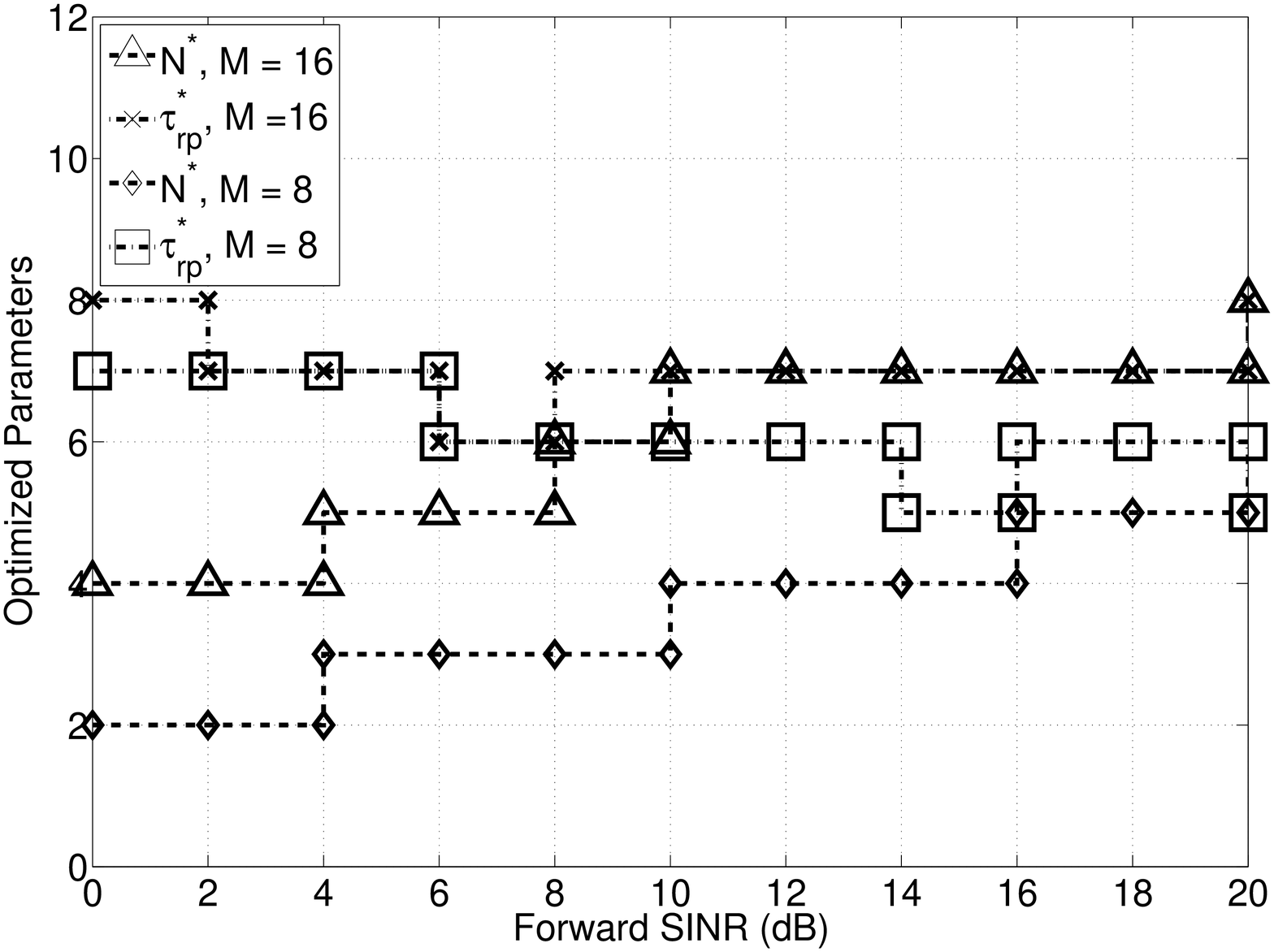}
\caption{Optimum values of parameters versus forward SINR}
\label{fig_sim4}
\end{figure}

In Fig. \ref{fig_sim4}, we plot the optimum values $\tau_{rp}^{*}$, $N^{*}$ and $K^{*}$, which maximize net throughput, versus forward SINR for $T=20$ symbols. Here, we fix the reverse SINR to be 10 dB less than the forward SINR. In all the cases plotted, the optimized value of the number of users $K^{*}=\tau_{rp}^{*}$. In Fig. \ref{fig_sim4}, we observe that the scheduling gains are more at low SINRs.

\subsection{Heterogeneous Users}

\begin{figure}[!t]
\centering
\includegraphics[width=88mm]{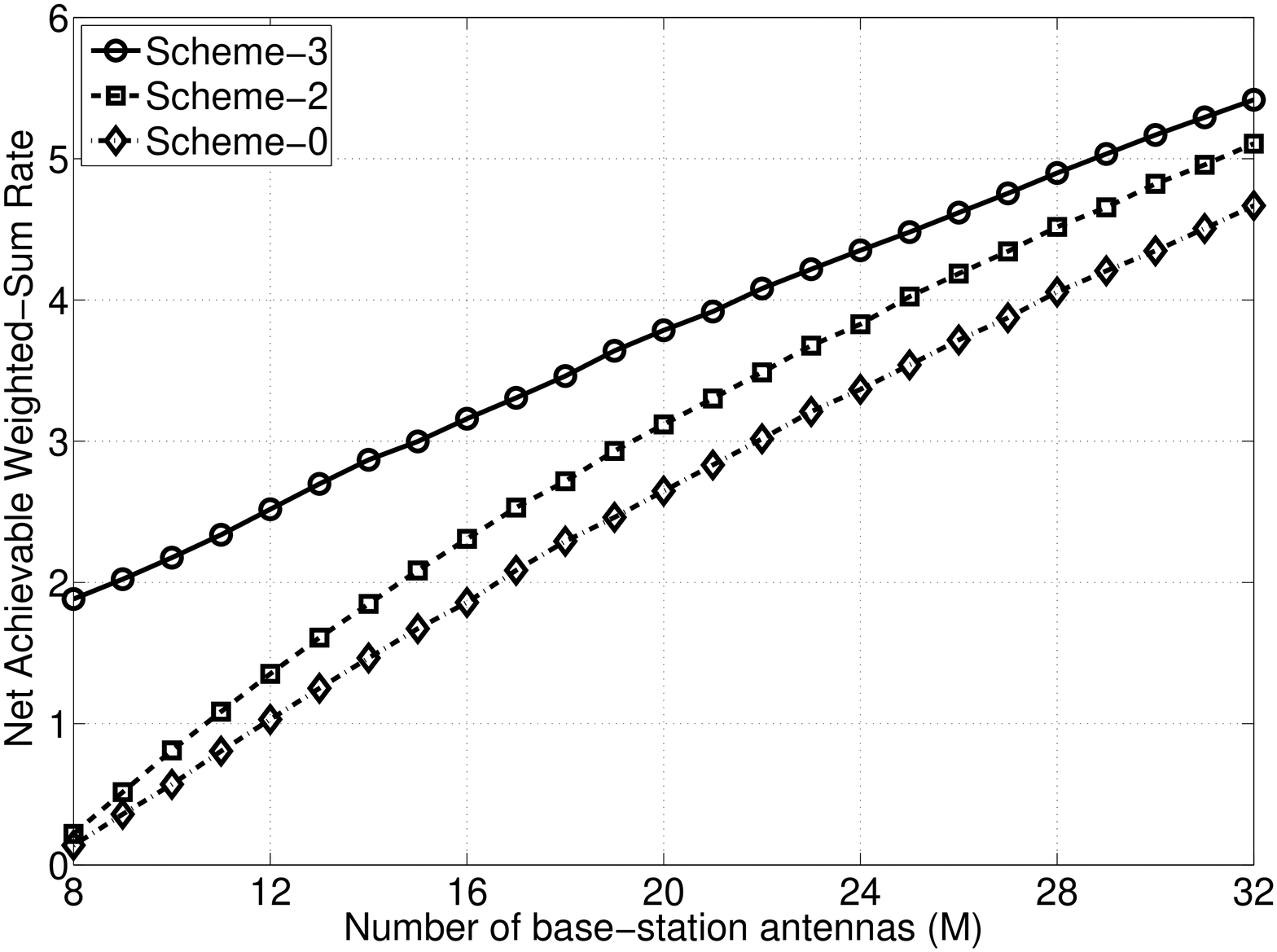}
\caption{Net achievable weighted-sum rate with optimized pre-conditioning (Scheme-2) and this combined with scheduling (Scheme-3)}
\label{fig_sim3}
\end{figure}

We consider a multi-user system consisting of $K=8$ users with forward SINRs $\{-4,-3,-2,-1,0,1,2,3\}$ dB and coherence interval $T=20$ symbols. We assume that the reverse SINR associated with every user is $10$ dB lower than its forward SINR. Next, we assign a weight of $2$ to the first four users and unit weight to the remaining users. The plot in Fig. \ref{fig_sim3} of net achievable weighted-sum rate versus $M$ clearly shows that using more antennas at the base-station is beneficial. Scheme-2 denotes optimized pre-conditioning with no scheduling and Scheme-3 denotes optimized pre-conditioning combined with scheduling. We observe that Scheme-3 gives the best performance. The performance gain due to scheduling is very significant when the number of users are comparable to the number of base-station antennas.

\section{Conclusion}
\label{concl}
\enlargethispage{-0.2in}

Our results show that even in interference-limited and highly mobile communication systems, the effective use of multiple antennas at the base-station greatly improve net downlink throughput in multi-user setting. We conclude that it is advantageous to increase the number of base-station antennas in the system model we considered. Reciprocal training made feasible by time-division duplex (TDD) operation is key to this result. With increase in the number of base-station antennas, the effective forward channel improves whereas the training sequence length required is not affected. The training sequence length has significant impact on the net throughput of mobile systems and hence, it is important to optimize it depending on various system parameters as discussed in the paper.

In multi-user multiple antenna systems, scheduling and pre-conditioning are practical schemes that can potentially improve the net throughput of these systems. We proposed scheduling schemes in both homogeneous and heterogeneous users scenarios and showed that these schemes significantly improve achievable sum/weighted-sum rate. The optimized pre-conditioning derived is applicable to the generic case with arbitrary set of weights, forward and reverse SINRs. Also, the optimization involved is computationally simple and can be implemented efficiently. As future work, we plan to extend these ideas to design a cellular network with aggressive frequency reuse supporting high mobility and high downlink rates.


\section*{Acknowledgment}

The authors would like to thank T. L. Marzetta for helpful discussions on this topic. This work was supported in part by NSF grants CNS-0626903 and CCF-0448181.





%


\end{document}